\documentclass{ws-procs}%\documentclass{ws-procs9x6square}
\usepackage{amsfonts}
\usepackage{epsfig}
\newcommand{\beq}{\begin{equation}}
\newcommand{\eeq}{\end{equation}}
\newcommand{\beqa}{\begin{eqnarray}}
\newcommand{\eeqa}{\end{eqnarray}}
\newcommand{\nn}{\nonumber \\}
\def \D {\Delta}
\def \Z {{\mathbb Z}}
\def \R {{\mathbb R}}
\def \PF {\mathrm{PF}}
\def \H {{\mathcal H}}
\def \Im {\mathrm{Im}\, }
\def \e {\mathrm{e}}

\def \s {\sigma}
\def \t {\tau}
\def \z {\zeta}
\def \ch {\mathrm{ch}}
\def \la {\langle}
\def \ra {\rangle}

\def \La {\underline{\Lambda}}

\begin{document}
%%%%%%%%%%%%%%%%%%%%%%%%%%%%%%%%%%%%%%%%%%%%%%%%%%%%%%%%%%%%%%
% title, author(s) and address(es) put here:                 %
%%%%%%%%%%%%%%%%%%%%%%%%%%%%%%%%%%%%%%%%%%%%%%%%%%%%%%%%%%%%%%

\title{The DIAGONAL AFFINE COSET CONSTRUCTION OF 
THE $\Z_k$ PARAFERMION HALL STATES}

%\author{A. Cappelli}
%\address{Istituto Nazionale di Fisica Nucleare, \\
%Via G. Sansone, 1,   50019 Sesto F.no (FI), Italy\\
%andrea.cappelli@fi.infn.it}

\author{L. Georgiev}
\address{Institute for Nuclear Research and Nuclear Energy\\
72 Tsarigradsko Chaussee, 1784 Sofia, Bulgaria\\
lgeorg@inrne.bas.bg}


\maketitle

\abstracts{
We construct  the $\Z_k$ parafermions as diagonal affine cosets
and apply them to the quantum Hall effect. This realization is 
particularly convenient for the analysis of the $\Z_k$ pairing rules,
the modular $S$-matrices, the $W_k$ symmetry and quantum group 
symmetry. The results are used for the computation of the mesoscopic 
chiral persistent currents in presence of Aharonov--Bohm flux.
}

%%%%%%%%%%%%%%%%%%%%%%%%%%%%%%%%%%%%%%%%%%%%%%%%%%%%%%%%%%%%%
% The main text of your paper                               %
%%%%%%%%%%%%%%%%%%%%%%%%%%%%%%%%%%%%%%%%%%%%%%%%%%%%%%%%%%%%%
%%%%%%%%%%%%%%%%%%%%%%%%%%%%%%%%%%%%%%%%%%%%%%%%%%%%%%%%%%%%
\section{Introduction}
Read and Rezayi \cite{rr} have described a new hierarchy of 
incompressible 
Hall states in the second Landau level with filling factors 
\beq\label{nu}
\nu_k= 2  +    \frac{k}{k+2}, \quad k=2,3,4 \ldots
\eeq
which have been observed later in an experiment \cite{pan} with 
thermal activation of quasiparticle--quasihole pairs 
in  extremely high mobility samples. The exact
quantization of the Hall conductance $\s_{H}=\nu (e^2/h) $ for 
filling factors $\nu=5/2$ and $8/3$  
($k=2$ and $4$) was obvious while at  $\nu= 13/5$ and $19/7$, 
($k=3$ and $5$) there were only well-expressed nonzero minimums of the 
resistance. 

The two-dimensional conformal field theory (CFT) has been successfully 
applied for 
the description of the dynamics of the edge excitations in the fractional 
quantum Hall (FQH) states. Due to the bulk incompressibility in the FQH 
sample the effective large-scale field theory turns out to be a
$(2+1)$ dimensional Chern--Simons theory \cite{fro} (say, on a dense 
cylinder) which is known to be 
equivalent to a $(1+1)$ dimensional  CFT (on a cylinder).
Thus, the universality classes of the incompressible FQH states can be 
ultimately labeled by unitary rational conformal field theories (RCFT).

The RCFT for the parafermion FQH states with filling factors~(\ref{nu}) 
has been originally formulated \cite{rr} in terms of a $u(1)$ boson 
describing the electric properties and the $\Z_k$ parafermionic CFT 
describing the neutral degrees of freedom. The ground state wave function
is realized as \cite{rr}
\beq\label{wave}
\Psi_{\mathrm{GS}}(z_1, \ldots, z_N)= 
\prod_{1\leq a<b\leq N} (z_a-z_b)^{1+\frac{2}{k}} \quad
\la \Psi_1(z_1) \cdots \Psi_1(z_N)\ra
\eeq 
where $\Psi_1$ is the first among the $\Z_k$-parafermion currents 
$\Psi_i$ for $i=1,\ldots, k-1$ with CFT dimensions $\D_i=i(k-i)/k$,  
and the following operator product expansion \cite{zf} 
\beqa\label{pf}
&&\Psi_i(z)\Psi_j(w)\sim \left(z-w \right)^{-\frac{2ij}{k}}  \Psi_{i+j}(w) ,
\quad i+j<k, \nn
&&\Psi_i(z)\Psi^*_j(w)\sim \left(z-w \right)^{-\frac{2j(k-i)}{k}}  
\Psi_{i-j}(w).
\quad i<j
\nonumber
\eeqa
In this paper we focus on an alternative realization of the $\Z_k$
parafermions in terms of a diagonal affine coset \cite{cgt2000} 
which turns out to be more natural for the parafermion FQH states.
%%%%%%%%%%%%%%%%%%%%%%%%%%%%%%%%%%%%%%%%%%%%%%%%%%%%%%%%%%%%%%%%%
\section{The  RCFT for the  $\Z_k$ parafermion  FQHE}
%%%%%%%%%%%%%%%%%%%%%%%%%%%%%%%%%%%%%%%%%%%%%%%%%%%%%%%%%%
The RCFT for the $\Z_k$ parafermion  FQH states has the same structure 
like the original one as formulated in Ref.~\cite{rr} 
\beq\label{PF}
\left(\widehat{u(1)} \oplus \PF_k \right)^{\Z_k},
\quad  
\PF_k=\frac{\widehat{su(k)_1}\oplus\widehat{su(k)_1}}{\widehat{su(k)_2}}.
\eeq
However, now the 
$\Z_k$ parafermions are realized as a diagonal affine coset
denoted as $\PF_k$ 
and we have used the superscript $\Z_k$ to denote a $\Z_k$ selection rule
coupling the electric and neutral sectors of the theory.
This rule can be formulated as follows: the edge excitations for the
parafermion FQH system are represented by certain primary conformal 
fields which are labeled by pairs of the $u(1)$ charge $\lambda$
and the parafermion primary fields $\Phi$
\[
(\lambda,\Phi) \ \Rightarrow \
:\e^{2\pi i \frac{\lambda}{\sqrt{k(k+2)}}\phi(z)}:  \otimes \  \Phi(z), \quad
\Phi(z) \in \PF_k^*.
\]
Then the $\Z_k$ paring rule (PR) states that an excitation labeled by 
$(\lambda,\Phi)$ is admissible only if 
\beq\label{PR}
P\left[ \Phi\right] = \lambda  \mod \ k ,
\eeq 
where $P$ is the parafermion $\Z_k$-charge \cite{zf,cgt2000}.
The topological order of the parafermion states, which is an important 
characteristics of any FQH state and gives basically the number of the 
topologically inequivalent excitations, in this case is \cite{cgt2000} 
\[
\mathrm{TO}= \frac{(k+1)(k+2)}{2}.
\]
It is instructive to note that the complete  parafermion CFT (\ref{PF})
can be obtained by projecting neutral degrees of freedom in a 
\textit{maximally symmetric chiral quantum Hall lattice} (in the 
terminology of Ref.~ \cite{fro}) denoted by the symbol 
$(3|{}^{1}A_{k-1}\, {}^{1}A_{k-1})$ that reproduces \cite{cgt2000} 
the same  filling factor (\ref{nu}).
The structure of this $K$-matrix theory corresponds to the left hand 
side of
\[
\left(\widehat{u(1)} \oplus
\widehat{su(k)_1}\oplus\widehat{su(k)_1} \right)^{\Z_k}\
\simeq \  \left( \widehat{u(1)} \oplus
\frac{\widehat{su(2)_k}}{\widehat{u(1)}} \oplus
\widehat{su(k)_2}  \right)^{\Z_k}
\]
while the $\widehat{su(k)_2} $ current algebra in the right hand side 
expresses the layer (or flavor)  symmetry of $k$ pairs of 
Dirac--Weyl fermions that could be gauged away without changing 
the filling factor since the subalgebra 
$\widehat{su(k)_1}\oplus \widehat{su(k)_1}$ is completely decoupled
from the electric charge. This projection provides a constructive way to 
obtain a non-abelian theory from an abelian parent 
(the latter being identified within the ADE classification of 
abelian FQH states \cite{fro}). 
It is worth-noting that the PR appears naturally \cite{cgt2000} in the 
abelian parent CFT and is preserved by the coset projection so that the 
projected theory inherits it.
%%%%%%%%%%%%%%%%%%%%%%%%%%%%%%%%%%%%%%%%%
\section{The diagonal coset $\PF_k$}
%%%%%%%%%%%%%%%%%%%%%%%%%%%%%%%%%%%%%%%
In this section we shall analyse in more detail the affine coset
\beq\label{PF_k}
\PF_k= \frac{\widehat{su(k)_1}\oplus \widehat{su(k)_1}}{\widehat{su(k)_2}}.
\eeq
The currents in the $\widehat{su(k)_2}$ algebra in the denominator of 
the coset are defined as 
\beq\label{J_2}
J^a=J^a_1+J^a_2 \in  \widehat{su(k)_2}, \quad \mathrm{where} 
\quad J^a_1 \in \widehat{su(k)_1}^{(1)},
\quad J^a_1 \in \widehat{su(k)_1}^{(2)}
\eeq
and the superscript $(1)$ or $(2)$ refers to the first  or the second copy of 
$\widehat{su(k)_1}$  in the numerator of $\PF_k$ respectively. 
That is why this type of cosets are called the ``diagonal'' cosets.

The stress tensor of the coset CFT (\ref{PF_k}) is constructed 
according to the standard 
Goddard--Kent--Olive procedure \cite{CFT-book}
\beq\label{T}
T(z)=
T^{\widehat{su(k)_1}\oplus \widehat{su(k)_1}}(z) -T^{\widehat{su(k)_2}}(z)
\eeq
and its modes satisfy the Virasoro algebra  with central charge 
\beq\label{c_k}
c_k= c^{(1)}+c^{(1)}-c^{(2)}=\frac{2(k-1)}{k+2},
\eeq
which is exactly the central charge of the $\Z_k$ parafermion CFT 
\cite{zf}.  Note that the 
stress tensor (\ref{T}) commutes with all currents in the 
$\widehat{su(k)_2}$ subalgebra which allows us to interpret the latter as 
a gauge symmetry.

\begin{figure}[htb]
\begin{center}
\epsfig{file=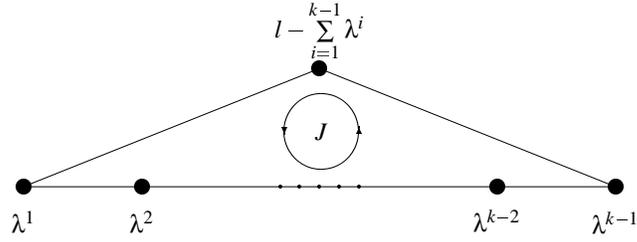,height=5cm,clip=,%
bbllx=140,bblly=380,bburx=515,bbury=560}
\vspace{-0.5cm}
\caption{Under the action of the simple current $J$ the Dynkin 
labels are rotated counter-clockwise, i.e.,  
$\lambda^1\to\lambda^2,\ldots,\lambda^{k-2}\to\lambda^{k-1}$, 
$\lambda^{k-1}\to\lambda^{0}$ and $\lambda^{0}\to\lambda^{1}$ 
\label{fig:dynkin}.}
\end{center}
\end{figure}

The primary fields of the coset $\PF_k$ can be labeled generically 
by triples
$(\La,\La';\La'')$ of SU(k) weights: two level-1 weights $\La$ and $\La'$ 
for the two copies of $\widehat{su(k)_1}$ in the numerator and one 
level-2 weight $\La''$ for the $\widehat{su(k)_2}$ in the denominator.
Not all combinations of weights are allowed \cite{schw}---they must
satisfy the following selection rule (conservation of $k$-ality)
\[
 [\La]+[\La']=[\La''], \quad \mathrm{where}\quad
[\La]=\sum_{i=1}^{k-1} i \lambda^i  \quad \mathrm{for}\quad
\La=\sum_{i=1}^{k-1} \lambda^i \La_i ,
\]
$\La_i$ are the fundamental weights for  $SU(k)$ and $\lambda^i$ 
the Dynkin labels corresponding to the general weight $\La$.
Since the admissible $\widehat{su(k)_1}$ weights are the fundamental 
weights $\La_\mu$ ($0\leq\mu\leq k-1$; $\La_0=0$) and  any admissible 
$\widehat{su(k)_2}$ weight  can be written as a 
sum of two fundamental weights, $\La_\mu+\La_\rho$, 
($0\leq\mu\leq\rho\leq k-1$)  
the admissible triples should have the form
\beq\label{admiss}
(\La_\mu,\ \La_\rho; \ \La_{\mu+\s} + \La_{\rho-\s}), 
\quad \mu,\rho,\s =0,\ldots, k-1.
\eeq
Next, not all admissible triples give inequivalent primary fields of 
the coset. The action of the simple currents \cite{schw}
\beq\label{simple}
J(\La_\mu)=\La_{(\mu+1\mod k)}
\eeq
which are outer automorphisms of the Dynkin diagrams, permuting the 
Dynkin labels as illustrated on Fig.~\ref{fig:dynkin}, 
commutes with the coset stress tensor while not commuting with
those of the numerator and denominator separately.
As a result the triples which can be connected by the action of the 
simple currents are equivalent and represent the same 
irreducible representation (IR) of the coset. This is known as the 
{\it field identification} \cite{schw}
\[
\left( J(\La),\ J(\La'); \ J(\La'') \right) \simeq 
\left( \La,\ \La' \ ; \ \La''  \right).
\]
Acting with the simple current (\ref{simple}) on the admissible triples
(\ref{admiss}) we get the following triple identification
\beq\label{ident}
(\La_{\mu+\gamma},\ \La_{\rho+\gamma}; \ \La_{\mu+\s+\gamma} +
\La_{\rho-\s+\gamma}) \simeq
(\La_{\mu},\ \La_{\rho}; \ \La_{\mu+\s} + \La_{\rho-\s}).
\eeq
Choosing $\gamma$ in the above relation such that $\rho+\gamma=0 \mod k$
one can always eliminate the second weight, which is our canonical choice 
of representative for the IR's of the coset. 
In this case the   $\widehat{su(k)_2}$-weight $\La''$ completely 
determines the triple since the first $\widehat{su(k)_1}$ weight is 
uniquely determined by its $k$-ality, which should match the $k$-ality
of the level-2 weight
\beq\label{IR}
\mathrm{\bf IR's:} \quad \La_{\mu} +  \La_{\rho}  \quad \Longleftrightarrow
\quad (\La_{\mu+\rho},\ \La_{0}; \ \La_{\mu} +  \La_{\rho}), \quad
0\leq\mu\leq\rho\leq k-1.
\eeq
Now the parafermion $\Z_k$ charge is simply 
\beq\label{P}
P[\La_{\mu} +  \La_{\rho}] = \mu +\rho \mod k.
\eeq
The definition of the coset stress tensor (\ref{T}) as the difference between 
two other stress tensors gives the following formula for the conformal 
dimensions
\beq\label{frac}
\D(\La,\La';\La'')=\D^{(1)}(\La)+ \D^{(1)}(\La') -  \D^{(2)}(\La'') \mod \Z
\eeq
which reproduces only the fractional part of the CFT dimension. 
Note that for some 
triples  the level-2 weight $\La''$ may have bigger positive CFT dimension 
than the sum of the CFT dimensions of $\La$ and $\La'$ in which case the 
total CFT dimension would appear to be negative, however, this is not 
the case. That is why ``$\mod \Z$'' in Eq.~(\ref{frac}) is necessary 
and important. 
Thus obtaining the correct CFT 
dimensions of the coset primary fields is usually indirect and non-trivial.
However, for this particular diagonal coset, using the field 
identification (\ref{ident}) and 
choosing appropriate representatives (different from the canonical ones)
  one can prove \cite{cgt2000}
the following exact formula for the CFT dimensions of the coset IR's
\beq\label{D}
\D^{\PF}(\La_\mu+\La_\rho)=\frac{\mu(k-\rho)}{k} +
\frac{(\rho-\mu)(k-\rho+\mu)}{2k(k+2)}, \quad 0\leq\mu\leq\rho\leq k-1.
\eeq
The characters of the diagonal coset 
\[
\ch_{\La_\mu +\La_\rho}(\t) =  
\mathop{\mathrm{tr}}_{\qquad\quad\H_{\La_\mu+\La_\rho} }
q^{L_0-c/24}, \quad q=\e^{2\pi i \t} \quad (\Im\t >0 )
\]
where $L_0$ is the zero modes of the stress tensor (\ref{T})
and $\H_{\La_\mu+\La_\rho}$ is the IR with label $\La_\mu+\La_\rho$, 
can be written in the form
of the so called \textit{universal chiral partition functions}  
with statistical matrix equal to the inverse Cartan matrix $C^{-1}$ 
for $SU(k)$
\beq\label{neut}
\ch^Q_{\s}(\t;\PF_k)= q^{ \D_\s - c_k/24 }
\sum\limits_{
\mathop{m_1,m_2,\ldots, m_{k-1}=0}\limits_{\sum\limits_{i=1}^{k-1}\, i
\, m_i \equiv Q \mod k } }^\infty
\frac{q^{\underline{m}. C^{-1}.
\left(\underline{m} - \La_\s \right)} }{(q)_{m_1} \cdots (q)_{m_{k-1}} },
\eeq
where $(q)_n=\prod\limits_{j=1}^n (1-q^j)$, $c_k$ is given by 
Eq.~(\ref{c_k}) and
$\D_\s = \D^{\PF}(\La_0+\La_\s)$ is computed from Eq.~(\ref{D}) for
$\s=0,\ldots, k-1$. The independent characters are labeled by
$(\s,Q)$ with $0\leq \s \leq Q \leq k-1$ and the relation between 
these new labels and the old ones is  $Q=\rho$, $\s = \rho-\mu$.
%%%%%%%%%%%%%%%%%%%%%%%%%%%%%%%%%%%%%%%%%%%%%%%%%%%%%%%%%%%%%%%%
\section{The modular $S$-matrix for the diagonal coset}
%%%%%%%%%%%%%%%%%%%%%%%%%%%%%%%%%%%%%%%%%%%%%%%%%%%%%%%%%%%%%%%
The $S$ transformation of the modular parameter $\t'=-1/\t$
(like in any RCFT) leads to the following linear transformation 
of the characters
\[  
\ch_{\La_\mu +\La_\nu}(\t')= \sum_{0\leq\rho\leq\s\leq k-1} 
{\mathop{S}\limits^{\circ}}_{\La_{\mu}+\La_{\nu};\ \La_{\rho}+\La_{\sigma}}
\quad\ch_{\La_\rho +\La_\s}(\t)
\]
where ${\mathop{S}\limits^{\circ}}$ is the coset $S$ matrix. 
Using the standard Goddard-Kent-Olive procedure for finding the 
$S$ matrix of the coset we obtain 
${\mathop{S}\limits^{\circ}}$
as a product  of the $S$ matrix of 
the numerator of the coset (\ref{PF_k}) and the inverse transposed 
$S$ matrix of the denominator \cite{CFT-book}. Since the numerator is a 
direct sum of algebras
 the $S$ matrices for both factors have to be multiplied, however, 
if we use only canonical triples (\ref{IR}) the second weight $\La'=0$ 
corresponds to the vacuum sector so that the second $S$ matrix becomes 
a simple normalization factor $1/\sqrt{k}$. The first $S$ matrix, which is of 
a rational $u(1)^{k-1}$ torus type, 
is completely determined from the $k$-alities 
of the labels  of $\widehat{su(k)_2}$ in the denominator.
Finally, since the $S$ matrix for $\widehat{su(k)_2}$  is unitary we get the 
following expression for the coset $S$ matrix:
\beq\label{Sc}
{\mathop{S}\limits^{\circ}}_{\La_{\mu}+\La_{\nu};\ \La_{\rho}+\La_{\sigma} }=
{ \exp\left(2\pi i \frac{(\mu+\nu)(\rho+\sigma)}{k}\right)} \
\left( { S^{(2)}_{\La_{\mu}+\La_{\nu};\ \La_{\rho}+\La_{\sigma} } }\right)^*
\eeq
where the label's parameters satisfy $\ 0\leq \mu\leq\nu\leq k-1$,
$0\leq\rho\leq\sigma \leq k-1$. 
Recall that the $S$ matrix for $\widehat{su(k)_2}$ can be computed using Kac 
formula \cite{CFT-book}
\[
S^{(2)}_{\La;\ \La'}=\frac{i^{\, k(k-1)/2}}{\sqrt{k(k+2)^{k-1}}}
\sum_{w\in \mathcal{W}} \epsilon(w) \
\exp\left( - 2\pi i
\frac{\left(\La+\underline{\rho}
\left|w(\La'+\underline{\rho})\right)\right.}{k+2}\right),
\]
where $\mathcal{W}$ is the Weyl group of $SU(k)$, $\epsilon(w)$ is 
the determinant of the element $w\in \mathcal{W}$ and 
$\underline{\rho}$ is the Weyl vector \cite{CFT-book}.

Given the explicit coset $S$ matrix (\ref{Sc}), the fusion rules can 
be obtained by the Verlinde formula \cite{CFT-book},
however, as can be seen  from Eq.~(\ref{Sc}) the coset $S$ matrix differs from
the (conjugated) $S$ matrix for $\widehat{su(k)_2}$ only by a phase. This phase
can be interpreted as the action of simple currents on $S^{(2)}$. 
Since the charge conjugation and the simple currents are automorphisms 
they preserve the fusion rules. Thus we conclude that the fusion rules for
 the diagonal coset  are the same as those for $\widehat{su(k)_2}$.
In particular there are primary fields which obey non-abelian statistics
\cite{cgt2000}.
%%%%%%%%%%%%%%%%%%%%%%%%%%%%%%%%%%%%%%%%%%%%%%%%%%%%%%%%%
\section{$W_k$ symmetry}
%%%%%%%%%%%%%%%%%%%%%%%%%%%%%%%%%%%%%%%%%%%%%%%%%%%%%%%%%%
The $\widehat{su(k)_2}$ currents which are gauged away in the 
diagonal coset (\ref{PF_k}) commute with all 
independent polynomial Casimir operators of $SU(k)$ which 
generate a $W_k$ algebra. Therefore, after removing all 
dimension-one currents, the coset (\ref{PF_k}) 
can be completely characterized by its $W_k$  symmetry \cite{cgt2000}.
The central charge (\ref{c_k}) of the diagonal coset coincides with 
that for the $W_k$ unitary minimal models with the lowest possible 
parameter $p=k+1$. To make this relation more transparent we shall 
identify the Fateev--Lukyanov  weight $\beta$ within the Coulomb gas 
approach to the $\Z_k$ parafermions \cite{zf} with the weights of the diagonal
 coset \cite{cgt2000}
\[
\beta =\alpha_+ \La^{(1)} + \alpha_- \La^{(2)}=
- \frac{k+1}{\sqrt{(k+1)(k+2)}} \left( \La_\mu +\La_{\rho}\right),
\] 
where, using canonical triples (\ref{IR}) only,   
we have set $\La^{(1)}=0$ and the Coulomb gas parameters are determined by
\[
\alpha_+ + \alpha_- = \alpha_0, \quad \alpha_+ \alpha_- =-1, \quad
(\alpha_0)^2=\frac{1}{(k+1)(k+2)}.
\] 
The explicit form of the $W_k$ generators can be given by using the 
quantum Miura transformation \cite{cgt2000}.
%%%%%%%%%%%%%%%%%%%%%%%%%%%%%%%%%%%%%%%%%%%%%%%%%%%%%%%%%%%%%%%%%%%
\section{Quantum group structure of $\PF_k$ }
%%%%%%%%%%%%%%%%%%%%%%%%%%%%%%%%%%%%%%%%%%%%%%%%%%%%%%%%%%%%%%%%%
The triples equivalence (\ref{ident})  and the canonical choice 
(\ref{IR}) demonstrate that there is a 1--1 correspondence between 
the IRs of $\PF_k$ and $\widehat{su(k)_2}$: both labeled by 
$\La_\mu+\La_\rho$ (this is not obvious in the 
$\widehat{su(2)_k}/\widehat{u(1)}$ realization of the $\Z_k$ 
parafermions). Furthermore, Eq.~(\ref{Sc}) proves that 
both  $S$-matrices are equivalent since they are related by  
charge conjugation and the action of simple currents. These two facts 
almost prove that the quantum group for the coset (\ref{PF_k})
is the same as that for $\widehat{su(k)_2}$, i.e., \cite{slin-bais}
\[
U_q\left( sl(k)\right) \quad \mathrm{with} \quad
q=\exp\left(-i\frac{\pi}{k+2}\right),  \quad
{\lfloor m\rfloor}_q=\frac{q^{m}-q^{-m}}{q^{1}-q^{-1}}.
\]
%%%%%%%%%%%%%%%%%%%%%%%%%%%%%%%%%%%%%%%%%%%%%%%%%%%%%%%%%%%%%%%%%%
\section{Application: persistent currents in the $\Z_k$ 
parafermion quantum Hall states}
\begin{figure}[htb]
\begin{center}
\epsfig{file=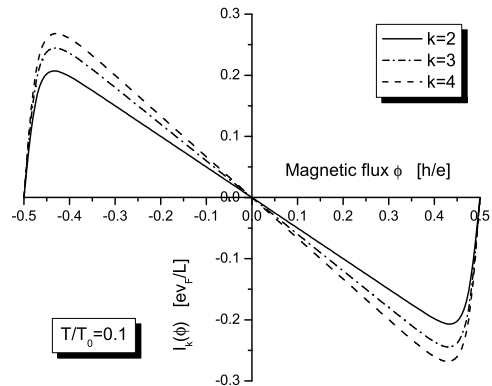,height=6cm,clip=,%
bbllx=0,bblly=0,bburx=270,bbury=207}
\caption{Persistent currents in the $k=2,3$ and $4$ parafermion FQH states,
as functions of the magnetic flux within one period,
computed numerically for $T/T_0=0.1$.
\label{fig:periods}}
\end{center}
\end{figure}

\begin{figure}[htb]
\begin{center}
\epsfig{file=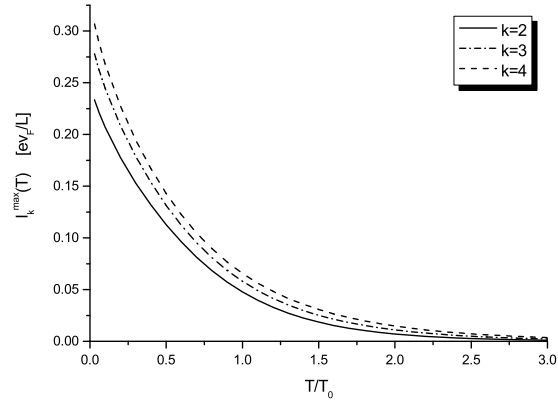,height=6cm,clip=,%
bbllx=4,bblly=0,bburx=292,bbury=220}
\caption{Temperature decay of the persistent current's amplitudes
in the $k=2,3$ and $4$ parafermion FQH states.
The zero temperature amplitudes in these units are $\nu_k/2$.
\label{fig:decay}}
\end{center}
\end{figure}

%%%%%%%%%%%%%%%%%%%%%%%%%%%%%%%%%%%%%%%%%%
When mesoscopic FQH disk samples are threaded by Aharonov--Bohm flux,  
the derivative of the free energy  $F(T,\phi)=-k_B T \ln Z(T,\phi)$, 
with respect to the magnetic flux $I(T,\phi)=- \partial_\phi F(T,\phi)$,
gives rise to  equilibrium persistent currents, which are observable
\[
I(T,\phi)=
\frac{e}{h}\, k_B T \frac{\partial}{\partial \phi}
\ln Z(\t,\phi\t), \quad 
Z(\t,\z) =  
\mathop{\mathrm{tr}}_{\quad\H }
q^{L_0-c/24}\ \e^{2\pi i \z J_0},
\]
whre $J_0$ is the electric charge operator.
The precise thermodynamic identification of the modular parameters
is given \cite{PRB-PF_k} in terms of the  absolute temperature $T$ and 
the (dimensionless)  Aharonov--Bohm flux $\phi=\Phi/(h/e)$
\[
q=\e^{2\pi i \t}, \quad \t=i\pi \frac{T_0}{T},\quad \z=\phi\t, \quad
T_0=\frac{\hbar v_F}{\pi k_B L}, \quad \phi\in \R.
\]
The complete chiral CFT partition function for the $\Z_k$ parafermion 
FQH states is constructed as the sum of the characters of all inequivalent IRs 
\cite{PRB-PF_k}
\[
Z_{k}(\t,\z) =
\e^{-{\pi}\nu_H\frac{\left(\Im\z\right)^2}{\Im\t}} 
\sum_{l  \mod  k+2} \quad
\sum_{\rho \geq l-\rho \mod k } \chi_{l,\rho}(\t,\z),
\]
where, following Ref.~\cite{cz}, a non-holomorphic exponential 
factor in front of the sum  has been included 
in order to restore the invariance of the FQH system under the 
Laughlin spectral flow. The complete IR's characters \cite{cgt2000}
\[
\chi_{l,\rho}(\t,\z) =
\sum_{s=0}^{k-1} K_{l+s(k+2)}(\t,k\z;k(k+2))
\ch\left(\La_{l-\rho+s} +\La_{\rho+s}\right)(\t)
\]
are special combinations (satisfying the $\Z_k$ pairing rule (\ref{PR}))
of the characters of the $u(1)$ sector
describing the electric/magnetic properties of the FQH system
\[
K_l(\tau,\zeta;m)=\frac{1}{\eta(\tau)}\sum_{n\in \Z}
q^{\frac{m}{2}(n+\frac{l}{m})^2} \
\mathrm{e}^{2\pi i \zeta (n+\frac{l}{m})},\quad
\eta(\t)=q^{\frac{1}{24}} \prod_{n=1}^\infty (1-q^n)
\]
and the neutral characters (\ref{neut}) of the diagonal coset $\PF_k$.
Using these explicit formulas we have computed numerically \cite{PRB-PF_k}
the chiral persistent currents for the $\Z_k$ parafermion FQH states
with $k=2,3$ and $4$
for temperatures in the range $0.03\leq T/T_0\leq 14$. 
As illustrated on Fig.~\ref{fig:periods} 
%%%%%%%%%%%%%%%%%%%%%%%
%%%%%%%%%%%%%%%%%%%%%% 
these 
currents are periodic functions of the AB flux with period exactly 
one quantum $h/e$ of flux. This is an important result since it forbids any 
opportunity for a spontaneous breaking of continuous symmetries 
\cite{PRB-PF_k}.
The amplitudes of these currents are universal and decay exponentially with 
increasing the temperature \cite{PRB-PF_k} as shown on Fig.~\ref{fig:decay}.
%%%%%%%%%%%%%%%%%%%%%%
%%%%%%%%%%%%%%%%%%%%%%
%%%%%%%%%%%%%%%%%%%%%%%%%%%%%%%%%%%%%%%%%%%%%%%%%%%%%%%%%%%%%
% Doing Acknowledgement                                     %
%%%%%%%%%%%%%%%%%%%%%%%%%%%%%%%%%%%%%%%%%%%%%%%%%%%%%%%%%%%%%
\section*{Acknowledgments}
Most of the talk followed the work \cite{cgt2000}.
The author thanks Christoph Schweigert for fruitful discussions,  
INFN--Firenze for hospitality and acknowledges partial support 
from the Bulgarian
National Council for Scientific Research under Contract F-828
as well as from the FP5-EUCLID Network Program  of the European 
Commission under Contract HPRN-CT-2002-00325.
%%%%%%%%%%%%%%%%%%%%%%%%%%%%%%%%%%%%%%%%%%%%%%%%%%%%%%%%%%%%%
% Doing Appendix(ices)                                      %
%%%%%%%%%%%%%%%%%%%%%%%%%%%%%%%%%%%%%%%%%%%%%%%%%%%%%%%%%%%%%
%
%%%%%%%%%%%%%%%%%%%%%%%%%%%%%%%%%%%%%%%%%%%%%%%%%%%%%%%%%%%%%
% Doing references:                                         %
%%%%%%%%%%%%%%%%%%%%%%%%%%%%%%%%%%%%%%%%%%%%%%%%%%%%%%%%%%%%%
%%%%%%%%%%%%%%%%%%%%%%%%%%%%%%%%%%%%%%%%%%%%% 
\def\NP{{\it Nucl. Phys.\ }} 
\def\PRL{{\it Phys. Rev. Lett.\ }} 
\def\PL{{\it Phys. Lett.\ }} 
\def\PR{{\it Phys. Rev.\ }} 
\def\CMP{{\it Comm. Math. Phys.\ }} 
\def\IJMP{{\it Int. J. Mod. Phys.\ }} 
\def\JSP{{\it J. Stat. Phys.\ }} 
\def\JP{{\it J. Phys.\ }} 
%%%%%%%%%%%%%%%%%%%%%%%%%%%%%%%%%%%%%%%%%%%%%% 

\end{document}